\documentclass[ALICE,manyauthors]{cernphprep}
\usepackage{hyperref}
\usepackage{lineno}
\usepackage[T1]{fontenc}
\usepackage[english]{babel}
\usepackage{graphicx}
\usepackage{verbatim}
\usepackage{amsfonts}
\usepackage{makeidx}
\usepackage{color}
\usepackage{amsmath}
\usepackage{lineno}
\usepackage{epsfig}
\usepackage{epstopdf}
\usepackage{hyperref}
\usepackage{cite}
\usepackage{textcomp}
\usepackage{multirow}
\usepackage{makecell}
\usepackage{booktabs}
\newcommand{\sqrtsNN}{\sqrt{s_{\rm \scriptscriptstyle NN}}}

\newcommand{\MeV}{\mathrm{MeV}}
\newcommand{\GeV}{\mathrm{GeV}}
\newcommand{\TeV}{\mathrm{TeV}}

\newcommand{\mum}{\mathrm{\mu m}}

\newcommand{\PbPb}{\mbox{Pb--Pb}}

\newcommand{\RaaFD}{R_{\rm AA}^\mathrm{feed\text{-}down}}

\newcommand{\pt}{p_{\rm T}}

\newcommand{\DtoKpi}{{\rm D}^0 \to {\rm K}^-\pi^+}
\newcommand{\DtoKpipi}{{\rm D}^+\to {\rm K}^-\pi^+\pi^+}
\newcommand{\DstartoDpi}{{\rm D}^{*+} \to {\rm D}^0 \pi^+}

\newcommand{\Dzero}{{\rm D^0}}

\newcommand{\Dstar}{{\rm D^{*+}}}

\newcommand{\Dplus}{{\rm D^+}}

\newcommand{\dEdx}{{\rm d}E/{\rm d}x}

\newcommand{\Jpsi}{{\rm J}/\psi}

\newcommand{\vn}{v_\mathrm{n}}
\newcommand{\vtwo}{v_\mathrm{2}}
\newcommand{\vthree}{v_\mathrm{3}}
\newcommand{\vntot}{v^\mathrm{tot}_\mathrm{n}}
\newcommand{\vnsig}{v^\mathrm{sig}_\mathrm{n}}
\newcommand{\vnbkg}{v^\mathrm{bkg}_\mathrm{n}}
\newcommand{\vnprompt}{v^\mathrm{prompt}_\mathrm{n}}
\newcommand{\fprompt}{f_\mathrm{prompt}}
\newcommand{\vnfd}{v^\mathrm{feed\text{-}down}_\mathrm{n}}
\newcommand{\Nsig}{N^\mathrm{sig}}
\newcommand{\Nbkg}{N^\mathrm{bkg}}
\newcommand{\Qn}{Q_\mathrm{n}}
\newcommand{\Rnflow}{R_{\rm n}}
\newcommand{\ppbar}{\mathrm{p+\bar{p}}}
\newcommand{\tamu}{TAMU}
\newcommand{\phsd}{PHSD}
\newcommand{\powlang}{POWLANG}
\newcommand{\powlanghtl}{POWLANG~HTL}
\newcommand{\powlanglqcd}{POWLANG~lQCD}
\newcommand{\bamps}{BAMPS}
\newcommand{\bampsel}{BAMPS$_{\rm el}$}
\newcommand{\bampselrad}{BAMPS$_{\rm el+rad}$}
\newcommand{\gossiaux}{MC@sHQ+EPOS2}
\newcommand{\lido}{LIDO}
\newcommand{\lbt}{LBT}
\newcommand{\lgr}{LGR}
\newcommand{\dabmod}{DAB-MOD}
\newcommand{\dabmodmet}{DAB-MOD(M\&T)}
\newcommand{\dabmodeloss}{DAB-MOD($E_{\rm loss}$)}
\newcommand{\catania}{Catania}

\newcommand{\pythiasixprecise}{PYTHIA 6.4.25}

\newcommand{\hijingprecise}{HIJING v1.383}
\newcommand{\fonll}{FONLL}
\newcommand{\evtgen}{EvtGen}

\graphicspath{{./}}

\begin{document}%

\begin{titlepage}
\PHyear{2020}
\PHnumber{082}      
\PHdate{19 May}  
%

\title{Transverse-momentum and event-shape dependence of D-meson flow harmonics in Pb--Pb collisions at $\pmb{\sqrtsNN = 5.02~\TeV}$}
\ShortTitle{$\pt$ and ESE dependence of D-meson $\vn$ harmonics}   

\Collaboration{ALICE Collaboration\thanks{See Appendix~\ref{app:collab} for the list of collaboration members}}
\ShortAuthor{ALICE Collaboration} 

\begin{abstract}
The elliptic and triangular flow coefficients $v_2$ and $v_3$ of prompt $\Dzero$, $\Dplus$, and $\Dstar$ mesons were measured at midrapidity ($|y|<0.8$) in Pb--Pb collisions at the centre-of-mass energy per nucleon pair of $\sqrtsNN=5.02~\TeV$ with the ALICE detector at the LHC. The D mesons were reconstructed via their hadronic decays in the transverse momentum interval $1<\pt<36~\GeV/c$ in central (0--10\%) and semi-central (30--50\%) collisions. Compared to pions, protons, and $\Jpsi$ mesons, the average D-meson $\vn$ harmonics are compatible within uncertainties with a mass hierarchy for $\pt\lesssim 3~\GeV/c$, and are similar to those of charged pions for higher $\pt$. The coupling of the charm quark to the light quarks in the underlying medium is further investigated with the application of the event-shape engineering (ESE) technique to the D-meson $v_2$ and $\pt$-differential yields. The D-meson $v_2$ is correlated with average bulk elliptic flow in both central and semi-central collisions. Within the current precision, the ratios of per-event D-meson yields in the ESE-selected and unbiased samples are found to be compatible with unity. All the measurements are found to be reasonably well described by theoretical calculations including the effects of charm-quark transport and the recombination of charm quarks with light quarks in a hydrodynamically expanding medium.
\end{abstract}
\end{titlepage}
\setcounter{page}{2}

\section{Introduction}

The formation of a strongly coupled colour-deconfined medium in ultra-relativistic heavy-ion collisions, called quark--gluon plasma (QGP), 
has been established both at RHIC and LHC energies~\cite{Muller:2006ee,Roland:2014jsa}.
The QGP behaves as a near-perfect fluid with small shear viscosity over entropy density ratio, $\eta/s$, undergoing an expansion that can be described by relativistic hydrodynamics~\cite{Heinz:2013th}.

In heavy-ion collisions, heavy quarks (charm and beauty) are predominantly produced via hard-scattering processes on a time scale shorter than the QGP formation time~\cite{Liu:2012ax,Andronic:2015wma}, and therefore they experience all the stages of the system evolution, interacting with the medium constituents via both elastic (collisional)~\cite{Braaten:1991we}  and inelastic (gluon radiation)~\cite{Gyulassy:1990ye,Baier:1996sk,Prino:2016cni} processes.
The measurement of the suppression of the yield of heavy-flavour hadrons in central nucleus--nucleus collisions relative to pp collisions scaled by the number of nucleon--nucleon collisions at both RHIC~\cite{Adare:2010de,Abelev:2006db,Adamczyk:2014uip,Adler:2005xv,Adare:2015hla} and LHC energies~\cite{Adam:2015sza,Abelev:2012qh,Adam:2016khe,Adam:2016wyz,Khachatryan:2016ypw,Sirunyan:2017xss,Acharya:2018hre} provides compelling evidence of heavy-quark energy loss in deconfined strongly interacting matter.

Additional insights into the QGP properties can be obtained by measuring the azimuthal anisotropy of heavy-flavour hadrons. In non-central nucleus--nucleus collisions the initial spatial anisotropy of the overlap region is converted via multiple interactions into an azimuthally anisotropic distribution in the momentum space of the produced particles~\cite{Ollitrault:1992bk,Voloshin:1994mz}. This anisotropy is characterised in terms of the Fourier coefficients $\vn=\langle \cos [\mathrm{n}(\varphi-\Psi_\mathrm{n})]\rangle$, where $\varphi$ is the azimuthal angle of the particle and $\Psi_\mathrm{n}$ is the azimuthal angle of the symmetry plane for the nth-order harmonic~\cite{Voloshin:1994mz,Poskanzer:1998yz}. The values of the Fourier coefficients depend on the geometry of the collision, the fluctuations in the distributions of nucleons and gluons within the nuclei~\cite{Qin:2010pf}, and the dynamics of the expansion. The second order flow coefficient $\vtwo$, called \textit{elliptic flow}, is related to the almond-shaped geometry of the overlap region between the colliding nuclei and, consequently, is the largest contribution to the anisotropy in non-central collisions. The third harmonic coefficient $\vthree$, named \textit{triangular flow}, originates from event-by-event fluctuations in the initial distribution of nucleons and gluons in the overlap region~\cite{Nahrgang:2014vza}. 
In particular, the measurement of the azimuthal anisotropy of heavy-flavour hadrons at low $\pt$ can help quantify the extent to which charm and beauty quarks participate in the collective expansion of the medium~\cite{Batsouli:2002qf}, as well as the fraction of heavy-flavour hadrons hadronising via recombination with flowing light quarks~\cite{Molnar:2004ph,Greco:2003vf}. At high $\pt$, instead, the charm hadron azimuthal anisotropy can constrain the path-length dependence of heavy-quark in-medium energy loss~\cite{Gyulassy:2000gk,Shuryak:2001me}. Precise measurements of heavy-flavour $\vn$ coefficients are useful to constrain the parameters of models that implement the heavy-quark transport in the QGP. In this context, the heavy-quark spatial diffusion coefficient $D_s$ in the QGP is particularly interesting, since it is related to the relaxation (equilibration) time of heavy quarks $\tau_\mathrm{Q}=(m_\mathrm{Q}/T)D_s$, where $m_\mathrm{Q}$ is the quark mass and $T$ is the medium temperature~\cite{Moore:2004tg}.

Further investigation into the dynamics of heavy quarks in the medium can be performed with the event-shape engineering (ESE) technique~\cite{Schukraft:2012ah}, which allows for selection of events with the same centrality but different initial geometry on the basis of the magnitude of the average bulk flow.
In fact, hydrodynamic calculations show that the average flow of the bulk of soft hadrons is proportional to the initial-state eccentricities~\cite{Beraudo:2018tpr} for small values of $\eta/s$~\cite{Heinz:2013th,Gardim:2012dc,Voloshin:2008dg}. By classifying the events with the ESE technique 
it is possible to investigate the correlation between the flow coefficients of D mesons and soft hadrons. According to the available calculations~\cite{Prado:2016szr,Gossiaux:2017nwz,Beraudo:2018tpr}, the initial system ellipticity is converted into parton flow with a similar efficiency for bulk and charm quarks, despite the different production mechanisms, dynamics, and hadronisation of heavy quarks and light partons forming the bulk of the medium. Moreover, the measurement of the D-meson spectra in events with different average eccentricity provides information about the possible correlation between the radial and elliptic flows at low-intermediate $\pt$, and the charm-quark energy loss and the elliptic flow at high $\pt$. The correlation with the radial flow is expected to be present from the observation of the scaling of the flow harmonics with the particle mass~\cite{Acharya:2018zuq}, while the correlation with the in-medium energy loss would be motivated by the different path traversed by the charm quark in the medium in the case of an isotropic or an eccentric system.

A positive D-meson $v_2$ is observed at both RHIC~\cite{Adare:2010de,Adamczyk:2014yew,Adamczyk:2017xur} and the LHC~\cite{Abelev:2013lca,Abelev:2014ipa,Adam:2016ssk,Adam:2015pga,Acharya:2017qps,Sirunyan:2017plt,Acharya:2018bxo}. The comparison of the D-meson $v_2$ with the charged-pion $v_2$ and with theoretical models~\cite{Uphoff:2012gb,He:2014cla,Monteno:2011gq,Cao:2013ita,Song:2015ykw,Nahrgang:2013xaa,Uphoff:2014hza,Beraudo:2014boa,Cao:2017hhk} indicates that charm quarks participate in the collective expansion of the medium and that both collisional processes and the recombination of charm and light quarks contribute to the observed elliptic flow. Furthermore, a positive $\Dzero$-meson $v_3$ was measured by the CMS Collaboration~\cite{Sirunyan:2017plt}.
The $\pt$-differential yields and $v_2$ of D mesons were measured by the ALICE Collaboration in samples of events selected on the basis of the average bulk elliptic flow with the ESE technique~\cite{Acharya:2018bxo}. A correlation between the D-meson $v_2$ and the $v_2$ of the bulk of light hadrons was observed, while the ratio of the $\pt$-differential yields in ESE-selected samples to the yields measured without any ESE selection was found to be compatible with unity within the large uncertainties.

In this Letter, the measurement of the non-strange D-meson flow harmonics performed on a large sample of Pb--Pb collisions at $\sqrtsNN=5.02$~TeV collected by ALICE in 2018 is reported. With this data sample, the D-meson $v_2$ is measured with the Scalar Product (SP) method in an extended $\pt$ interval and with smaller uncertainties with respect to the previous results obtained with the Event Plane (EP) method described in~\cite{Acharya:2017qps, Acharya:2018bxo} in the 30--50\% centrality class. The results obtained with the SP and the EP method were found to be compatible between each other, as reported in previous publications~\cite{Abelev:2014ipa}. The measurement of the D-meson $v_2$ coefficient in the 0--10\% centrality class and $v_3$ coefficient in the 0--10\% and 30--50\% centrality classes are also presented. In addition, the measurement of the $v_2$ and the modification of the $\pt$ distributions in the ESE-selected samples is reported in narrower classes of the average event flow with respect to~\cite{Acharya:2018bxo}. The measurements are compared to theoretical calculations in order to assess information about the participation of the charm quark in the collective motion of the system and its interactions with the QGP constituents.

\section{Detector and data sample}
\label{sec:detector}
A detailed description of the ALICE apparatus and data acquisition framework can be found in~\cite{Aamodt:2008zz,Abelev:2014ffa}. The main detectors used for this analysis are the Inner Tracking System (ITS)~\cite{Aamodt:2010aa}, the Time Projection Chamber (TPC)~\cite{Alme:2010ke}, and the Time-Of-Flight (TOF) detector~\cite{Akindinov:2013tea}. The ITS is a six-layer silicon detector which provides the event selection, the reconstruction of primary and secondary vertices, and the tracking of charged particles. The TPC detector is used for the track reconstruction and the particle identification (PID) via the measurement of the specific energy loss $\dEdx$, while the TOF detector provides PID via the measurement of the flight time of the particles. These detectors are located inside a solenoid providing a uniform magnetic field of 0.5 T parallel to the LHC beam direction and cover the pseudorapidity interval $|\eta| <$ 0.9. 
A minimum-bias interaction trigger was provided by the coincidence of signals in the two scintillator arrays of the V0 detector~\cite{Abbas:2013taa}, covering the full azimuth in the pseudorapidity regions $-3.7 < \eta < -1.7$ (V0C) and $2.8 < \eta < 5.1$ (V0A). An online selection based on the V0 signal amplitudes was applied in order to enhance the sample of central and mid-central collisions through two separate trigger classes. Background events from beam--gas interactions were removed offline using the time information provided by the V0 and the neutron Zero-Degree Calorimeters (ZDC)~\cite{Arnaldi:1999zz}. Only events with a primary vertex reconstructed within $\pm 10$~cm from the centre of the detector along the beam line were considered in the analysis.

Events were divided into centrality classes, defined in terms of percentiles of the hadronic Pb--Pb cross section, using the amplitudes of the signals in the V0 arrays.
The number of events in each centrality class considered for this analysis (0--10\% and 30--50\%) is about $100\times 10^6$ and $85\times 10^6$, corresponding to an integrated luminosity of $\simeq 130~\mu \mathrm{b}^{-1}$ and $\simeq56~\mu\mathrm{b}^{-1}$, respectively~\cite{PhysRevC.97.054910}.
In order to apply the ESE technique, the events in each centrality class were further divided into samples with different average elliptic anisotropy of final-state particles, selected according to the magnitude of the second-order harmonic reduced flow vector $q_{2}$~\cite{Voloshin:2008dg}, defined as
		\begin{equation}
		q_2 = |\pmb{Q}_2| / \sqrt{M},	
		\label{eq:q2}
		\end{equation}
		where $M$ is the number of tracks used in the $|\pmb{Q}_2|$ calculation selected as described below, and 	
		\begin{equation}
		\pmb{Q}_2 = \sum_{k=1}^{M} e^{i2\varphi_k}
		\label{eq:QvecTPC}
		\end{equation}
		is a vector built from the azimuthal angles ($\varphi_k$) of the considered particles.
The $\pmb{Q}_2$ vector was measured using charged tracks reconstructed in the TPC with $|\eta|<0.8$ and $0.2<\pt<5~\GeV/c$ to exploit the $\varphi$ resolution of the TPC and the large multiplicity at midrapidity, which are crucial to maximise the selectivity of $q_2$ with respect to the final state flow eccentricity~\cite{Adam:2015eta,Acharya:2018bxo}. The denominator in Eq.~\ref{eq:q2} is introduced to remove the dependence of $|\pmb{Q}_2|$ on $\sqrt{M}$ in the absence of flow~\cite{Voloshin:2008dg}. The tracks used to form the D-meson candidates were excluded from the computation of $q_2$ to partially remove autocorrelations between D mesons and $q_2$. 
The effect of residual autocorrelations between the D mesons and $q_2$ was studied in~\cite{Acharya:2018bxo} by computing $q_2$ from the azimuthal distribution of the energy deposition measured in the V0A detector, and hence introducing a pseudorapidity gap of two units between the D mesons and $q_2$. The $\vtwo$ values obtained with the $q_2$ calculated with TPC tracks and using the V0 detector were found to be compatible with a reduction of the eccentricity discriminating power of the two detectors without allowing for a firm conclusion on the magnitude of non-flow contamination. The same study was repeated for the data sample used for this analysis, leading to the same conclusions.

The selection of the events according to the average elliptic flow of the event was performed by defining $q_2$ percentiles in 1\%-wide centrality intervals as described in~\cite{Acharya:2018bxo} and~\cite{ALICE-PUBLIC-2020-003} to avoid the introduction of biases in the centrality (multiplicity) distribution of the selected events. The ESE-selected classes defined for the analyses presented in this paper correspond to the 20\% of events with smallest and largest $q_2$, respectively, and will be indicated as ``small-$q_2$'' and ``large-$q_2$''. In case of no ESE selection, the term ``unbiased'' will be used.

\section{Analysis technique}
\label{sec:analysis}
The charmed mesons were reconstructed at midrapidity via the decay channels $\DtoKpi$ (with branching ratio, $\mathrm{BR}= 3.89\pm0.04\%$), $\DtoKpipi$ ($\mathrm{BR} = 8.98\pm0.28\%$), and $\DstartoDpi$ ($\mathrm{BR}=67.7\pm0.5\%$) and their charge conjugates \cite{Tanabashi:2018oca}. $\Dzero$ and $\Dplus$ candidates were built combining pairs and triplets of tracks with the proper charge, $\pt>0.4~\GeV/c$, $|\eta|<$ 0.8, a fit quality \mbox{$\chi^2$/ndf $<$ 2} in the TPC (where ndf is the number of degrees of freedom involved in the track fit procedure), at least 70 (out of 159) associated space points in the TPC, and a minimum number of two hits in the ITS, with at least one in the two innermost layers.
$\Dstar$ candidates were formed by combining $\Dzero$ candidates with low-$\pt$ tracks, referred to here as ``soft pions'', which were required to have $\pt>$ 0.1 $\GeV/c$, $|\eta|<$ 0.8, and at least three associated hits in the ITS.
These selections limit the D-meson acceptance in rapidity, which drops to zero for $|y|>0.6$ for $\pt=1~\GeV/c$ and $|y|>0.8$ for $\pt>5~\GeV/c$. A $\pt$-dependent fiducial acceptance cut, $|y_{\mathrm{D}}| < y_{\rm fid}(\pt)$, was therefore applied, defined as a second-order polynomial function increasing from 0.6 to 0.8 in the range $1<\pt<5~\GeV/c$, and fixed to 0.8 for $\pt>5~\GeV/c$.

The D-meson candidate selection approach adopted to reduce the combinatorial background is similar to that used in previous analyses~\cite{Abelev:2014ipa, Acharya:2017qps}. The analysis procedure searches for decay vertices displaced from the primary vertex, exploiting the mean proper decay lengths of about 123 and 312 $\mum$ for $\Dzero$ and $\Dplus$ mesons, respectively~\cite{Tanabashi:2018oca}. The variables mainly used to distinguish between signal and background candidates are based on the separation between the primary and decay vertices, the displacement of the tracks from the primary vertex, and the pointing angle of the reconstructed D-meson momentum to the primary vertex, and are the same as described in~\cite{Acharya:2017jgo, Acharya:2018hre}.
In the strong decay of the $\Dstar$ meson the primary vertex cannot be differentiated from the secondary vertex. Therefore the geometrical selections were applied on the secondary vertex topology of the produced $\Dzero$ mesons.
The optimisation of the selection criteria for each D-meson species was performed as a function of $\pt$ and independently for the two centrality classes. 
Further reduction of the combinatorial background was obtained by applying PID for the daughter tracks with the TPC and TOF detectors. A selection in units of resolution ($\pm 3\,\sigma$) was applied on the difference between the measured and expected signals of pions and kaons for both $\dEdx$ and time-of-flight. The same selections are applied both for the unbiased and the ESE-selected measurements.

The D-meson elliptic and triangular flow coefficients, $v_2$ and $v_3$, were measured using the Scalar Product (SP) method~\cite{Voloshin:2008dg, Luzum:2012da, Adler:2002pu}. For each D-meson candidate, the $\vn$ coefficients can be expressed in terms of the $\Qn$ vectors, introduced in Sec.~\ref{sec:detector}, as
\begin{equation}
\label{eq:SP}
\vn\{\mathrm{SP}\}=\langle\langle \pmb{u}_{\mathrm{n}}\cdot\frac{\pmb{Q}_{\mathrm{n}}^{\rm A*}}{M^{\rm A}}\rangle\rangle\bigg{/}\sqrt{\frac{\langle\frac{\pmb{Q}_{\mathrm{n}}^{\rm A}}{M^{\rm A}}\cdot\frac{\pmb{Q}_{\mathrm{n}}^{\rm B*}}{M^{\rm B}}\rangle\langle\frac{\pmb{Q}_{\mathrm{n}}^{\rm A}}{M^{\rm A}}\cdot\frac{\pmb{Q}_{\mathrm{n}}^{\rm C*}}{M^{\rm C}}\rangle}{\langle\frac{\pmb{Q}_{\mathrm{n}}^{\rm B}}{M^{\rm B}}\cdot\frac{\pmb{Q}_{\mathrm{n}}^{\rm C*}}{M^{\rm C}}\rangle}},
\end{equation} 
where \textbf{\textit{u}}$_{\mathrm{n}} = e^{i\mathrm{n}\varphi_\mathrm{D}}$ is the unit flow vector of the D-meson candidate with azimuthal angle $\varphi_{\mathrm{D}}$ and the symbol "*" denotes the complex conjugation. The denominator is called the resolution ($\Rnflow$) and is calculated with the formula introduced in~\cite{Voloshin:2008dg}, where the three subevents, indicated as A, B, and C, are defined by the particles measured in the V0C, V0A, and TPC detectors, respectively. $\pmb{Q}^k_{\mathrm{n}}$ is the subevent flow vector corresponding to the nth-order harmonic for the subevent $k$, and $M^k$ represents the subevent multiplicity. This is defined as the sum of the amplitudes measured in each channel for the V0A and the V0C. 
For the V0A and V0C detectors, the $\Qn$ vectors were calculated from the azimuthal distribution of the energy deposition, and their components are given by
\begin{equation}
	Q_{\mathrm{n},x}^\text{V0A or V0C} = \sum_{k=1}^{N_{\rm sectors}} w_k\cos (\mathrm{n}\varphi_k),\quad Q_{\mathrm{n},y}^\text{V0A or V0C} = \sum_{k=1}^{N_{\rm sectors}} w_k\sin (\mathrm{n}\varphi_k),
	\label{eq:QvecV0}
\end{equation}
where the sum runs over the 32 sectors ($N_{\rm sectors}$) of the V0A or V0C detector, $\varphi_k$ is the azimuthal angle of the centre of the sector $k$, and $w_k$ is the amplitude measured in sector $k$, once the gain of the single channels is equalised and the recentering is applied to correct effects of non-uniform acceptance~\cite{Selyuzhenkov:2007zi}. For the TPC detector, the $\Qn$ vectors were computed using Eq.~\ref{eq:QvecTPC}.
The single bracket $\langle\rangle$ in Eq.~\ref{eq:SP} refers to an average over all the events, while the double brackets $\langle\langle\rangle\rangle$ denote the average over all particles in the considered $\pt$ interval and all events. The $\Rnflow$ is obtained as a function of the collision centrality.

\begin{figure*}[!t]
\begin{center}
\includegraphics[width=1.\textwidth]{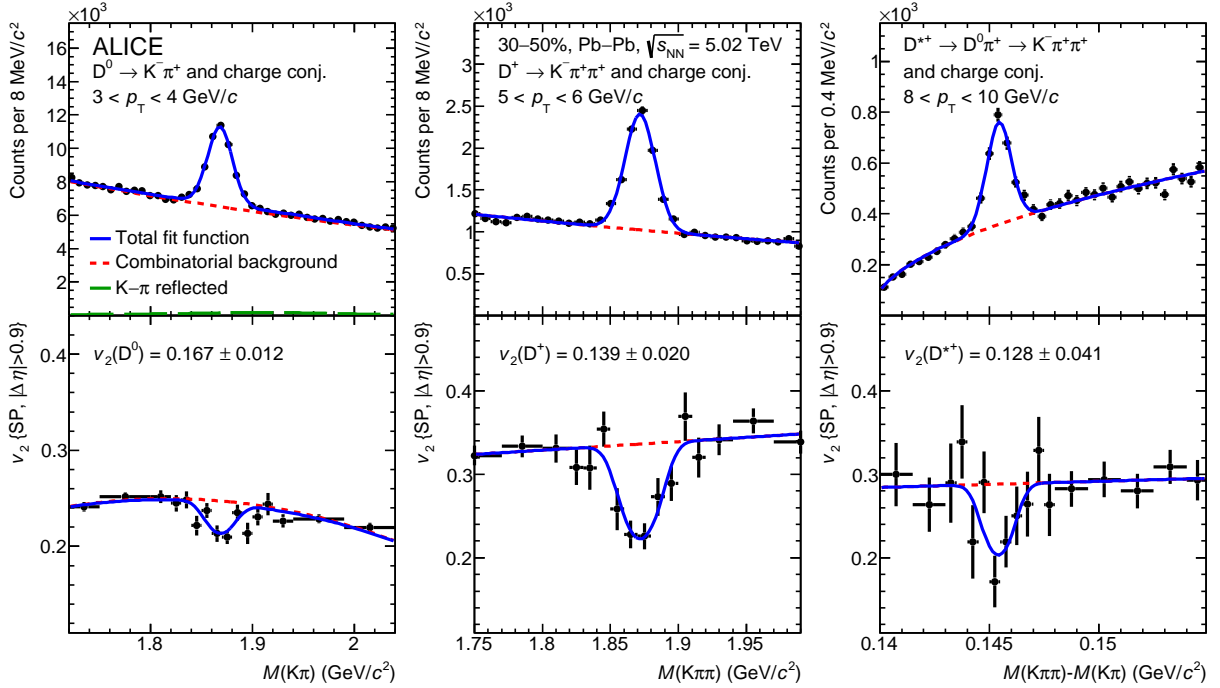}
\caption{Simultaneous fits to the invariant-mass spectrum and $v_2$ ($M_{\rm{D}}$) of $\Dzero$ (left panel), $\Dplus$ (middle panel), and $\Dstar$ (right panel) meson candidates in the $3<\pt<4~\GeV/c$, $5<\pt<6~\GeV/c$, and $8<\pt<10~\GeV/c$ intervals, respectively, for the 30--50\% centrality class. The solid blue and the dotted red curves represent the total and the combinatorial-background fit functions, respectively. For the $\Dzero$ candidates, the green dashed curve represents the contribution of the reflected signal.} 
\label{fig:Dmass_vs_v2} 
\end{center}
\end{figure*}

The $\vn$ of the D mesons cannot be directly measured using Eq.~\ref{eq:SP} as $\Dzero$, $\Dplus$, and $\Dstar$ cannot be identified on a particle-by-particle basis. Therefore, a simultaneous fit to the invariant-mass spectrum and the $\vntot$ distribution as a function of the invariant mass ($M_\mathrm{D}$) was performed in each $\pt$ interval for $\Dzero$ and $\Dplus$ candidates separately in order to measure the raw yields and the flow coefficients. For the $\Dstar$ case the distributions are studied as a function of the mass difference $\Delta M = M(\mathrm{K}\pi\pi) - (\mathrm{K}\pi)$. The measured anisotropic flow coefficient, $\vntot$, can be written as a weighted sum of the $\vn$ of the D-meson candidate, $\vnsig$, and that of background, $\vnbkg$~\cite{Borghini:2004ra} as
\begin{equation}
\label{eq:vnTot}
	\vntot (M_\mathrm{D}) = \vnsig \frac{\Nsig}{\Nsig + \Nbkg}(M_\mathrm{D}) + \vnbkg (M_\mathrm{D})\frac{\Nbkg}{\Nsig + \Nbkg}(M_\mathrm{D}),
\end{equation}
where $\Nsig$ and $\Nbkg$ are the raw signal and background yields, respectively. The fit function for the invariant-mass distributions was composed of a Gaussian term to describe the signal and an exponential distribution for the background for $\Dzero$ and $\Dplus$ candidates, while for the $\Dstar$ candidates the background was described by the function $a\sqrt{\Delta M - m_{\pi}}e^{b(\Delta M - m_{\pi})}$, where $a$ and $b$ are free parameters. In the case of the $\Dzero$ invariant mass the contribution of signal candidates with the reflected K-$\pi$ mass assignment was taken into account with an additional term. Its invariant-mass distribution was parameterised with a double-Gaussian distribution based on Monte Carlo (MC) simulations~\cite{Acharya:2019mgn,Acharya:2018bxo,Acharya:2017jgo,Acharya:2017qps,Abelev:2014ipa}. In the MC simulation, the underlying Pb--Pb events at $\sqrtsNN = 5.02~\TeV$ were simulated using the \hijingprecise{} generator~\cite{Wang:1991hta} and $\rm c\overline c$ or $\rm b\overline b$ pairs were added with the \pythiasixprecise{} generator~\cite{Sjostrand:2006za} with Perugia-2011 tune~\cite{Skands:2010ak}. 
In the simultaneous fit, the $\vn$ parameter for the candidates with wrong K-$\pi$ mass assignment was set to be equal to $\vnsig$, provided that the origin of these candidates are real $\Dzero$ mesons. 
The $\vnsig$ is measured from the fit to the $\vntot$ distribution with the function of Eq.~\ref{eq:vnTot}, where $\vnbkg$ is a linear function for $\Dplus$ and $\Dstar$ mesons, and $\Dzero$ mesons with $\pt>4~\GeV/c$. For the $\Dzero$ candidates with $\pt<4~\GeV/c$, a second-order polynomial function was used instead. 
Figure \ref{fig:Dmass_vs_v2} shows the simultaneous fit to the invariant-mass spectrum and $v_2^\mathrm{tot}(M_\mathrm{D})$ in the $\pt$ intervals 3--4 $\GeV/c$ for $\Dzero$, 5--6 $\GeV/c$ for $\Dplus$, and 8--10 $\GeV/c$ for $\Dstar$ in the 30--50\% centrality class.

The reconstructed D-meson signal is a mixture of prompt D mesons from $c$-quark hadronisation or strong decays of excited charmonium or open-charm states, and D mesons from beauty-hadron decays, called ``feed-down" in the following. The $\vnsig$ is therefore a linear combination of prompt ($\vnprompt$) and feed-down ($\vnfd$) contributions, and can be expressed as
\begin{equation}
\label{eq:vnprompt}
\vnsig = \fprompt \vnprompt + (1 - \fprompt) \vnfd,
\end{equation}

where $\fprompt$ is the fraction of promptly produced D mesons estimated as a function of $\pt$ with the theory-driven method described in~\cite{Acharya:2018hre}. This method uses (i) \fonll{} calculations~\cite{Cacciari:1998it,Cacciari:2001td} for the production cross section of beauty hadrons, (ii) the beauty-hadron decay kinematics from the \evtgen{} package~\cite{Lange:2001uf}, (iii) the product of efficiency and acceptance $({\rm Acc}\times\varepsilon)$ from Monte Carlo simulations, and (iv) a hypothesis on the nuclear modification factor of feed-down D mesons. 

The anisotropic flow coefficients of promptly produced D mesons were obtained assuming $\vnfd=\vnprompt/2$. The hypothesis is based on the measurement of the non-prompt $\Jpsi$ performed by CMS \cite{Khachatryan:2016ypw} and on the available model calculations \cite{Uphoff:2012gb,Aichelin:2012ww,Greco:2007sz}, that indicate $0< \vnfd < \vnprompt$.

For the measurement of the modification of the $\pt$-differential distributions of D mesons in the ESE-selected samples compared to the unbiased sample, the raw yields were extracted via fits to the invariant-mass distributions of $\Dzero$, $\Dplus$, and $\Dstar$ candidates and normalised to the corresponding number of events in the corresponding ESE-selected sample. The same functions adopted in the simultaneous fits for the invariant-mass distributions were used. The extracted raw yields were not corrected for the efficiency in the ratio calculation, under the assumption that the reconstruction and selection efficiencies do not depend on $q_2$. This assumption was previously verified in~\cite{Acharya:2018bxo}. 

\section{Systematic uncertainties}
The D-meson $\vn$ coefficients are affected by the systematic uncertainties due to (i) the signal extraction from the invariant-mass and $\vntot$ distributions, (ii) the beauty feed-down contribution, (iii) on the selection of the centrality interval in which $\Rnflow$ is calculated, and (iv), for the ESE-selected samples, the uncertainties due to possible bias caused by the presence of auto-correlation effects between the subevents used for $\Rnflow$ and $q_2$ calculations.

The uncertainty due to the simultaneous fit was estimated by repeating the fit several times with different configurations. In particular, the lower and upper limits of the fit range, the bin width, and the background fit functions used for the invariant-mass and $\vntot$ distributions were varied. For each configuration the D-meson $\vn$ was calculated and the absolute systematic uncertainty for each $\pt$ interval was assigned as the r.m.s. of the $\vn$ distribution obtained from the different trials.
The absolute systematic uncertainty values on the $\vn$ are reported in Table~\ref{tab:syst} and they depend on the D-meson species, the $\pt$ interval and the ESE-selected class. This source of uncertainty was considered as uncorrelated among the $\pt$ intervals and the centrality classes for the two harmonics.
The correlation between the small-$q_2$/large-$q_2$ and the unbiased case was investigated and the outcome indicated that this uncertainty source is uncorrelated between the different $q_2$-selected samples.

For the $\pt$-differential yield ratios in ESE-selected samples, the uncertainty for the signal extraction was estimated using the same approach described above, directly on the ratio of the yields in the ESE-selected and unbiased samples, leading to a systematic uncertainty value from 0.7$\%$ to 5$\%$, depending on the $\pt$ and the D-meson species.

\begin{table}[!tb]
\caption{Summary of systematic uncertainties on the measurement of the D-meson $\vtwo$, in the unbiased and ESE-selected samples, and $\vthree$ in Pb--Pb collisions at $\sqrtsNN=5.02~\TeV$. The range of the uncertainties on the fitting procedure and feed-down subtraction are quoted as absolute uncertainties, while those on the $\Rnflow$ as relative uncertainty.}
\centering
\renewcommand*{\arraystretch}{1.2}
\begin{tabular}[t]{l|>{\centering}p{0.14\linewidth}>{\centering}p{0.14\linewidth}>{\centering}p{0.14\linewidth}>{\centering\arraybackslash}p{0.14\linewidth}}
\toprule
Systematic uncertainty source & $\vtwo$ & $\vthree$ & $\vtwo$ small-$q_2$ & $\vtwo$ large-$q_2$\\
\midrule
\midrule
 & \multicolumn{4}{c}{0--10\%} \\
\midrule
$M$ and $\vn$ fits & 0.005--0.03 & 0.006--0.03 & 0.006--0.01 & 0.006--0.01 \\
 Feed-down & 0.002--0.01 & 0.0007--0.01 & 0.0003--0.006 & 0.003--0.016 \\ 
 $\Rnflow$ determination & 3.5$\%$ & negl. & 3.5$\%$ & 3.5$\%$ \\
  Autocorrelations on $R_{\rm 2}$ and $q_2$ & - & - & 3.5$\%$ & 1$\%$ \\
\midrule
\midrule
 & \multicolumn{4}{c}{30--50\%} \\
\midrule
$M$ and $\vn$ fits & 0.006--0.025 & 0.01--0.05 & 0.006--0.015 & 0.004--0.015 \\
 Feed-down & 0.0004--0.02 & 0.003--0.018 & 0.003--0.01 & 0.004--0.029 \\
 $\Rnflow$ determination & 0.5$\%$ & 0.5$\%$ & 0.5$\%$ & 0.5$\%$ \\
 Autocorrelations on $R_{\rm 2}$ and $q_2$& - & - & 0.5$\%$ & 0.5$\%$ \\
\bottomrule
\end{tabular}
\label{tab:syst}	
\end{table}

The systematic uncertainty source related to the beauty feed-down correction has two main contributions. The first is due to the $\fprompt$ calculation and it was studied by varying the quark mass and the renormalisation and factorisation scales in the \fonll{} calculations, the $\RaaFD$ hypothesis as reported in~\cite{Acharya:2018hre}. The second contribution is due to the assumption of $\vnfd$ = $\vnprompt$/2, previously described in Sec.~\ref{sec:analysis}, and was estimated by assuming a flat distribution of $\vnfd$ between 0 and $\vnprompt$ and by varying the central value of $\vnfd$ by $\pm\vnprompt$/$\sqrt{12}$. The values of the absolute systematic uncertainty from the beauty feed-down correction are reported in Table~\ref{tab:syst} and they depend on the D-meson species, the $\pt$ interval and the ESE-selected class. The uncertainty due to the beauty feed-down correction was assumed to be fully correlated among the $\pt$ bins for the measured $\vn$ coefficients in the same centrality class.

The non-flow effects are naturally suppressed because of the pseudorapidity gap of at least 0.9 units between the pseudorapidity interval used for the D-meson reconstruction, and the V0C used for the $\Qn$-vector determination. Furthermore, the auto-correlation effect due to the usage of the TPC tracks for the $q_2$ estimate has been discussed in Sec.~\ref{sec:detector} and the related systematic uncertainty was found to be negligible, as described in~\cite{Acharya:2018bxo}.

The contribution of the $\Rnflow$ to the systematic uncertainty is due to the centrality dependence. The central value of $\Rnflow$ was estimated using the three subevent formula, as described in Sec.~\ref{sec:analysis}, averaged over the events in the 0--10\% and 30--50\% intervals. The uncertainty was evaluated as the difference of the centrality integrated $\Rnflow$ values with those obtained as weighted averages of $\Rnflow$ values in narrow centrality intervals using the D-meson yields as weights. A systematic uncertainty of 3.5$\%$ and 0.5$\%$ was assigned on $R_{\rm 2}$ in the 0--10\% and 30--50\% centrality classes and for all ESE-selected samples. For the $R_3$, an uncertainty of 0.5$\%$ was assigned in the 30--50\% interval while it was found to be negligible for the 0--10\% class. The uncertainty associated with the resolution factor is smaller for the third harmonic than for the second harmonic, due to the milder centrality dependence of $R_3$ compared with that of $R_2$.

For the ESE-selected samples an additional source of systematic uncertainty on the resolution originates from auto-correlations due to the usage of the TPC tracks both for $q_2$ and $R_{\rm 2}$ determination. This potential bias is assessed by replacing the ratio $\langle\pmb{Q}_{\mathrm{n}}^{\rm V0C}/M^{\rm V0C}\cdot\pmb{Q}_{\mathrm{n}}^{\rm TPC*}/M^{\rm TPC}\rangle$/$\langle\pmb{Q}_{\mathrm{n}}^{\rm V0A}/M^{\rm V0A}\cdot\pmb{Q}_{\mathrm{n}}^{\rm TPC*}/M^{\rm TPC}\rangle$ in Eq.~\ref{eq:SP} with the one from the $q_2$-integrated analysis, following the same approach used for the $\Jpsi$ azimuthal anisotropy measurement \cite{Acharya:2018pjd}. In this case, the systematic uncertainty was estimated to be $3.5\%$ for the small-$q_2$ and $1\%$ for the large-$q_2$ samples in the 0--10\% centrality class, and $0.5\%$ for both $q_2$-selected classes in the 30--50\% centrality class, as reported in Table~\ref{tab:syst}. The last two sources of systematic uncertainty, related to the resolution, are considered to be fully correlated among the different $\pt$ intervals.

For the analysis of the $\pt$-differential yield ratios in ESE-selected and unbiased samples the reconstruction efficiency was verified to be independent of $q_2$. Consequently, it cancels out in the ratio of the two ESE-selected classes.
\section{Results}
\subsection{Unbiased flow harmonics}
\label{sec:results_unb}
\begin{figure*}[!t]
\begin{center}
\includegraphics[width=0.9\textwidth]{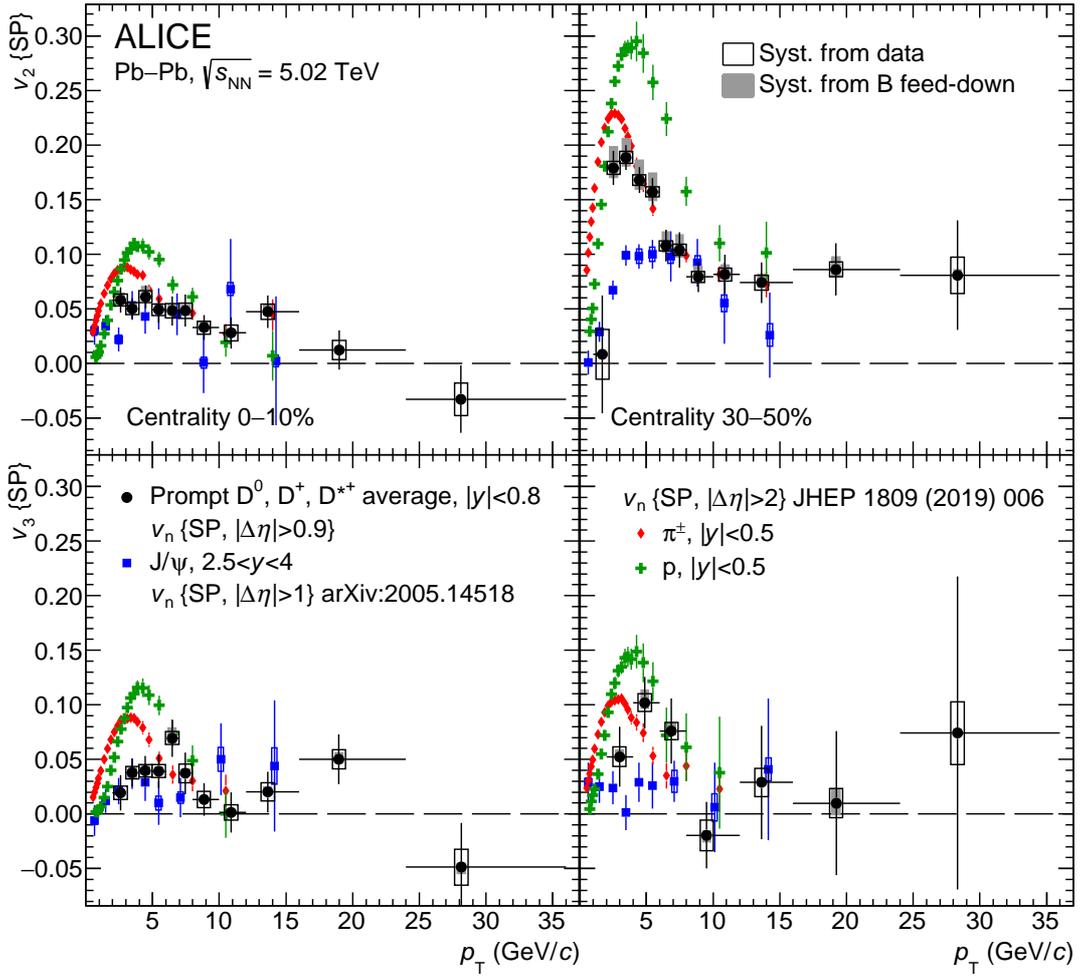}
\caption{Average $v_2$ (top panels) and $v_3$ (bottom panels) coefficients of prompt $\Dzero$, $\Dplus$, and $\Dstar$ mesons as a function of $\pt$ for $\PbPb$ collisions at $\sqrtsNN=5.02~\TeV$ in the 0--10\% (left panels) and 30--50\% (right panels) centrality classes. The $v_2$ and $v_3$ of $\pi^\pm$, $\ppbar$~\cite{Acharya:2018zuq} and inclusive $\Jpsi$ mesons~\cite{Acharya:2020jil} measured at the same centre-of-mass energy and in the same centrality classes are shown for comparison.} 
\label{fig:Daverage_v2_v3} 
\end{center}
\end{figure*}

Figure~\ref{fig:Daverage_v2_v3} shows the average $v_2$ (top panels) and $v_3$ (bottom panels) coefficients of prompt $\Dzero$, $\Dplus$, and $\Dstar$ mesons measured in the unbiased sample as a function of $\pt$ in the 0--10\% (left panels) and 30--50\% (right panels) centrality classes. The average $\vn$ of prompt $\Dzero$, $\Dplus$, and $\Dstar$ mesons was computed by using the inverse squared absolute statistical uncertainties as weights, after having compared their compatibility~\cite{ALICE-PUBLIC-2020-003}. The systematic uncertainties were propagated to the average by considering the contributions from the centrality dependence of the $\Rnflow$ resolution and the correction for the beauty feed-down component in the D-meson yields as correlated among the D-meson species. The D-meson $\vn$ harmonics are compared to the corresponding coefficients measured for charged pions and protons at midrapidity ($|y|<0.5$)~\cite{Acharya:2018zuq} as well as to inclusive $\Jpsi$ mesons at forward rapidity ($2.5<y<4$)~\cite{Acharya:2020jil}.

The D-meson elliptic flow increases significantly from central to semi-central collisions, as expected from the increasing eccentricity of the interaction region. Conversely, the triangular flow is compatible in the two centrality classes within the large uncertainties, following the milder centrality dependence of the third flow harmonic observed for light-flavour particles~\cite{Acharya:2018zuq}. For $\pt<3\text{--}4~\GeV/c$ ($\pt<4\text{--}5~\GeV/c$) the measured D-meson $v_2$ ($v_3$) is lower than that of pions and protons. This observation is consistent within uncertainties with the hypothesis of a mass hierarchy, $\vn(\mathrm{D}) < \vn(\mathrm{p}) < \vn(\pi)$, in the low $\pt$ region ($\pt\lesssim 3~\GeV/c$). In semi-central events the $\vn$ coefficients of $\Jpsi$ mesons seem to follow the mass hierarchy ($\vn(\Jpsi)<\vn(\mathrm{D})$). In central events the data suggests a similar behaviour, however within the current uncertainties no firm conclusions can be drawn. This observation can be explained by the interplay between the anisotropic flow and the isotropic expansion of the system (radial flow), which imposes an equal velocity boost to all particles. For $4\lesssim\pt\lesssim 6\text{--}8~\GeV/c$, the D-meson $\vn$ coefficients are similar to those of charged pions and lower than those of protons. This observation is consistent with a scaling of the $\vn$ coefficients with the number of constituent quarks, which supports the hypothesis of particle production via quark coalescence~\cite{Molnar:2003ff}. In the same $\pt$ interval, for the 30--50\% centrality class the larger values of $\vn$ for D mesons compared to $\Jpsi$ mesons can be explained by (i) the hadronisation via coalescence together with the larger flow coefficients of up and down quarks compared to that of charm quarks~\cite{Molnar:2004ph} and (ii) the fraction of $\Jpsi$ mesons coming from beauty-hadron decays~\cite{Adam:2015rba, Chatrchyan:2012np}, which are expected to have lower $v_2$ and $v_3$ than charmed mesons~\cite{Beraudo:2017gxw, Nahrgang:2014vza}. In the 0--10\% centrality class, the current experimental uncertainties do not allow for firm conclusions on the expected difference for $\Jpsi$ and D mesons. The measured $\vn$ coefficients for all the hadron species are compatible within uncertainties for $\pt\gtrsim 8~\GeV/c$. Similar values of $\vn$ coefficients are expected, because in this kinematic range the charm-quark mass is small compared to the momentum, and because the path-length dependence of the in-medium parton energy loss is similar for high-$\pt$ charm quarks and gluons.
\begin{figure*}[!t]
\begin{center}
\includegraphics[width=0.9\textwidth]{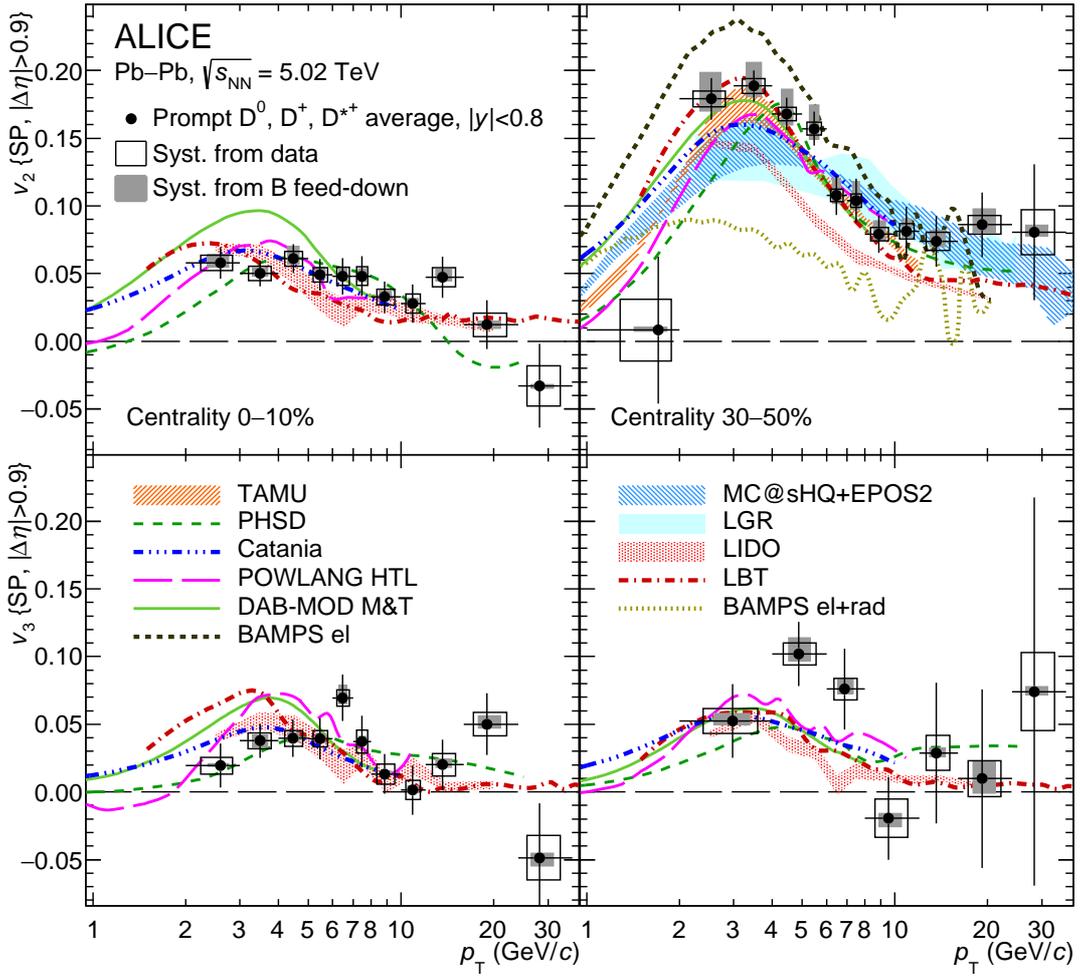}
\caption{Average $v_2$ (top panels) and $v_3$ (bottom panels) coefficients of prompt $\Dzero$, $\Dplus$, and $\Dstar$ mesons as a function of $\pt$ for $\PbPb$ collisions at $\sqrtsNN=5.02~\TeV$ in the 0--10\% (left panels) and 30--50\% (right panels) centrality classes compared with model calculations~\cite{He:2019vgs,Beraudo:2014boa,Beraudo:2018tpr,Uphoff:2012gb,Uphoff:2014hza,Song:2015ykw,Nahrgang:2014vza,Cao:2016gvr,Cao:2017hhk,Ke:2018tsh,Ke:2018jem,Prado:2016szr,Katz:2019fkc,Scardina:2017ipo,Plumari:2019hzp}.} 
\label{fig:Daverage_v2_v3_vsModels} 
\end{center}
\end{figure*}

In Fig.~\ref{fig:Daverage_v2_v3_vsModels}, the average D-meson $\vn$ coefficients are compared to theoretical calculations that include the charm-quark transport in a hydrodynamically expanding medium. The theoretical uncertainties, where available, are displayed with a coloured band. In the \tamu~\cite{He:2019vgs}, \powlanghtl~\cite{Beraudo:2014boa,Beraudo:2018tpr}, \phsd~\cite{Song:2015ykw}, \catania~\cite{Scardina:2017ipo,Plumari:2019hzp}, and \bampsel~\cite{Uphoff:2012gb} calculations the interactions between the charm quarks and the medium constituents are modelled with collisional processes, while the \gossiaux~\cite{Nahrgang:2014vza}, \lbt~\cite{Cao:2016gvr,Cao:2017hhk}, \lido~\cite{Ke:2018tsh,Ke:2018jem}, \bampselrad~\cite{Uphoff:2014hza}, \dabmodmet~\cite{Prado:2016szr,Katz:2019fkc}, and \lgr~\cite{Li:2019lex} models include also radiative processes. The difference in the variants of the \bamps{} model indicates that in this model elastic collisions are the dominant process that imparts a positive D-meson $v_2$ in the low and intermediate $\pt$ region. All the models except for \bamps{} include the hadronisation of the charm quark via coalescence, in addition to the fragmentation mechanism. Initial-state event-by-event fluctuations are included in the \powlanghtl, \lido, \phsd, \gossiaux, \lbt, and \dabmodmet{} models, which are therefore the only ones that provide predictions for the triangular flow.
Although the models differ in several aspects related to the interactions both in the QGP and in the hadronic phase as well as to the medium expansion, most of them provide a fair description of the measured $\vn$ harmonics. The largest difference is observed in the $2<\pt<6~\GeV/c$ interval for the $v_2$ in the 30--50\% centrality class, where most of the models provide a prediction lower than the measured points. This is more evident for the \lido{} model, which shows a deviation of $5.4\,\sigma$, and \bampselrad{}, which underestimates the measured $v_2$ by about a factor two with more than $10\,\sigma$ significance. In contrast to this, \bampsel{} overestimates the measurement by about $3\,\sigma$. The underestimation of the data by the \bampselrad{} model can be eventually due to the missing implementation of the charm-quark coalescence with light quarks from the medium, which seems to be necessary in the description of the measured $v_2$. In the same $\pt$ range, the \dabmod{} model overestimates the measured $v_2$ in the 0--10\% centrality class by $3.7\,\sigma$. These discrepancies expressed in number of standard deviations were computed combining the probability to observe a deviation from the null hypothesis (i.e. the model prediction) for all the measured points in the $2<\pt<6~\GeV/c$ interval, considering both the experimental (statistical and systematic) and the theoretical uncertainties, when available.
\begin{table}[!tb]
\caption{Summary of $\chi^2/\mathrm{ndf}$ values obtained for the different model predictions compared with the measured D-meson $\vn$ harmonics.}
\centering
\renewcommand*{\arraystretch}{1.2}
\begin{tabular}[t]{l|>{\centering}p{0.12\linewidth}>{\centering}p{0.081\linewidth}>{\centering}p{0.081\linewidth}>{\centering}p{0.0001\linewidth}>{\centering}p{0.081\linewidth}>{\centering}p{0.081\linewidth}>{\centering\arraybackslash}p{0.08\linewidth}}
\toprule
\multirow{3}[4]{*}{Model} & \multirow{3}[4]{*}{$\pt$ ($\GeV/c)$} & \multicolumn{6}{c}{$\chi^2/\mathrm{ndf}$} \\
\cline{3-8}
& &  \multicolumn{2}{c}{$\vtwo$} &  & \multicolumn{2}{c}{$\vthree$} & \multirow{2}[3]{*}{global}\\
\cline{3-4}
\cline{6-7}
& &  0--10\% & 30--50\% & & 0--10\% & 30--50\% & \\
\midrule
\midrule
\bampsel{}~\cite{Uphoff:2012gb} & [1--24] & - & 31.7/11 && - & - & -\\
\bampselrad{}~\cite{Uphoff:2014hza} & [1--24] & - & 203.6/11 && - & - & -\\
\catania{}~\cite{Scardina:2017ipo,Plumari:2019hzp} & [1--12] & 3.1/7 & 14.0/8 && 15.1/7 & 8.1/4 & 40.3/26\\
\dabmodmet{}~\cite{Prado:2016szr,Katz:2019fkc} & [1--8] & 24.6/7 & 9.8/6 && 16.1/7 & 7.1/3 & 57.6/23 \\
\lbt{}~\cite{Cao:2016gvr,Cao:2017hhk} & [1--36] & 18.2/11 & 15.8/12 && 24.9/11 & 8.4/7 & 67.4/41\\
\lido{}~\cite{Ke:2018tsh,Ke:2018jem} & [1--24] & 10.7/10 & 62.0/11 && 17.8/10 & 12.5/6 & 102.9/37\\
\lgr{}~\cite{Li:2019lex} & [1--24] & - & 15.5/11 && - & - & -\\
\gossiaux{}~\cite{Nahrgang:2014vza} & [1--36] & - & 15.7/12 && - & - & -\\
\phsd{}~\cite{Song:2015ykw} & [1--24] & 13.2/10 & 19.6/11 && 7.9/10 & 8.6/6 & 48.9/37\\
\powlanghtl{}~\cite{Beraudo:2014boa,Beraudo:2018tpr} & [1--12] & 9.6/7 & 13.5/8 && 14.6/7 & 8.3/4 & 45.9/26\\
\tamu{}~\cite{He:2019vgs}& [1--12] & - & 8.15/9 && - & - & -\\

\bottomrule
\end{tabular}
\label{tab:chi2}	
\end{table}
The global agreement between the data and the theoretical models was evaluated by computing the $\chi^2/\mathrm{ndf}$, as done in~\cite{Acharya:2017qps}. The values are reported in Table~\ref{tab:chi2}. All the centrality classes and $\vn$ harmonics were considered when the model predictions were available. Compared to the results in~\cite{Acharya:2017qps}, for almost all the models the $\chi^2/\mathrm{ndf}$ is found to be higher than unity, most likely because of the improved precision of the measurement. The models that describe the data with $\chi^2/\mathrm{ndf}<2$ are \gossiaux, \lbt, \lgr, \phsd, \powlang, \catania, and \tamu{} which is more in agreement with the data compared to~\cite{Acharya:2017qps}, thanks to the improved description of the charm-quark coalescence in its latest version~\cite{He:2019vgs}. These models use a value of heavy-quark spatial diffusion coefficient in the range $1.5<2\pi D_s T_\mathrm{c}<7$ at the critical temperature $T_\mathrm{c} = 155~\MeV$~\cite{Borsanyi:2010bp}, which is consistent with the interval obtained in~\cite{Acharya:2017qps}. It is however important to consider that not all the theoretical models provide predictions for all the $\vn$ harmonics in all the centrality classes reported in this article, hence the global interpretation of these comparisons could not be conclusive.

\subsection{Event-shape engineered flow harmonics and $\pmb{\pt}$-differential yields}
\begin{figure*}[t]
\begin{center}
\includegraphics[width=1.\textwidth]{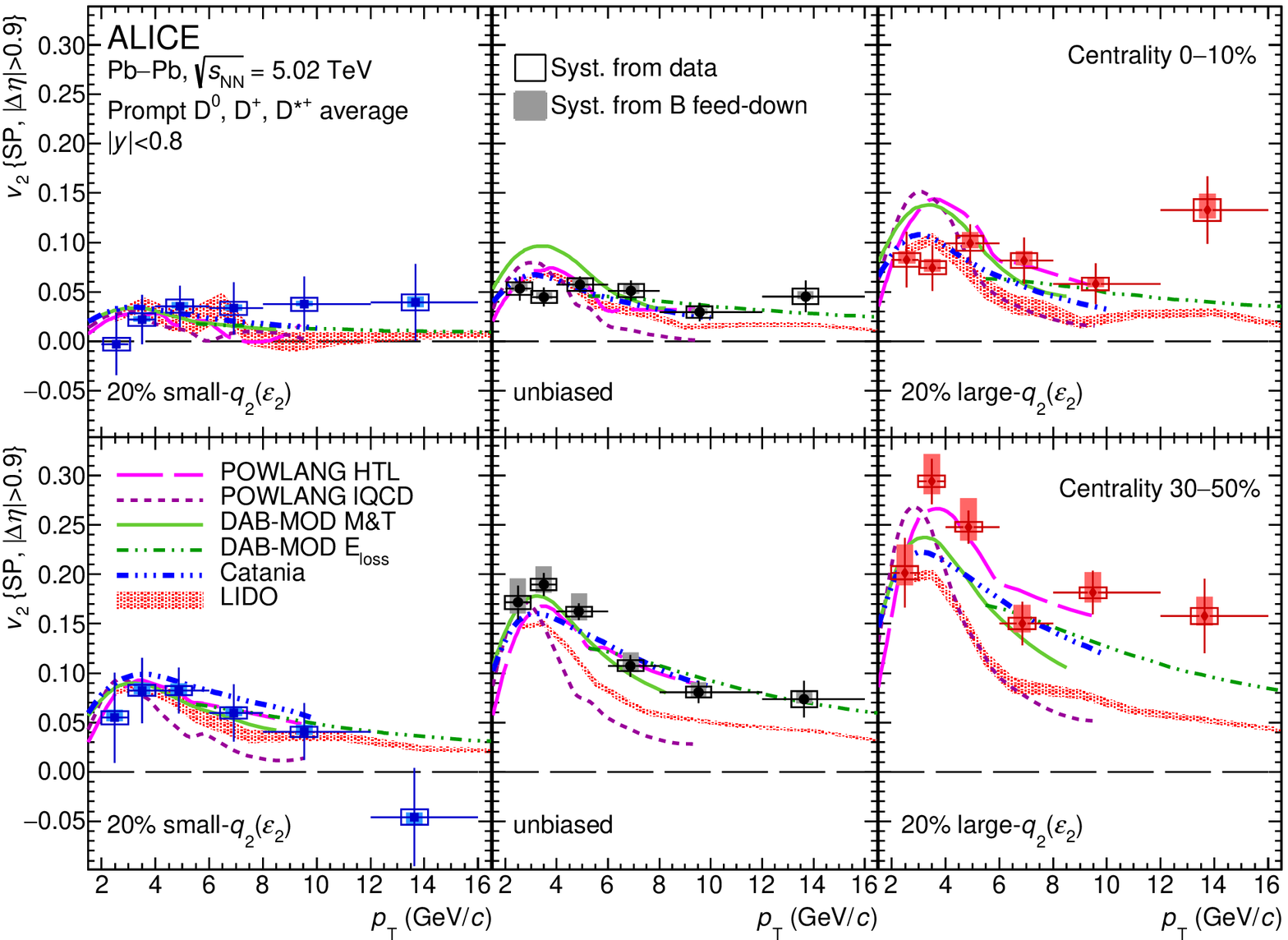}
\caption{Average of prompt $\Dzero$, $\Dplus$, and $\Dstar$ meson $v_2$ as a function of $\pt$ in $\PbPb$ collisions at $\sqrtsNN=5.02~\TeV$ in the small-$q_2$, large-$q_2$ (see text for details), and unbiased samples, for the 0--10\% (top panels) and 30--50\% (bottom panels) centrality classes, compared to model calculations~\cite{Beraudo:2018tpr,Ke:2018tsh,Prado:2016szr,Katz:2019fkc,Scardina:2017ipo,Plumari:2019hzp}. In the \lido{}, \dabmod{}, and \catania{} predictions, the ESE selection is performed with a $q_2$ estimator, while in the \powlang{} model the elliptic eccentricity $\epsilon_2$ is used.} 
\label{fig:Daverage_v2_ESE} 
\end{center}
\end{figure*}

\begin{figure*}[!t]
\begin{center}
\includegraphics[width=1.\textwidth]{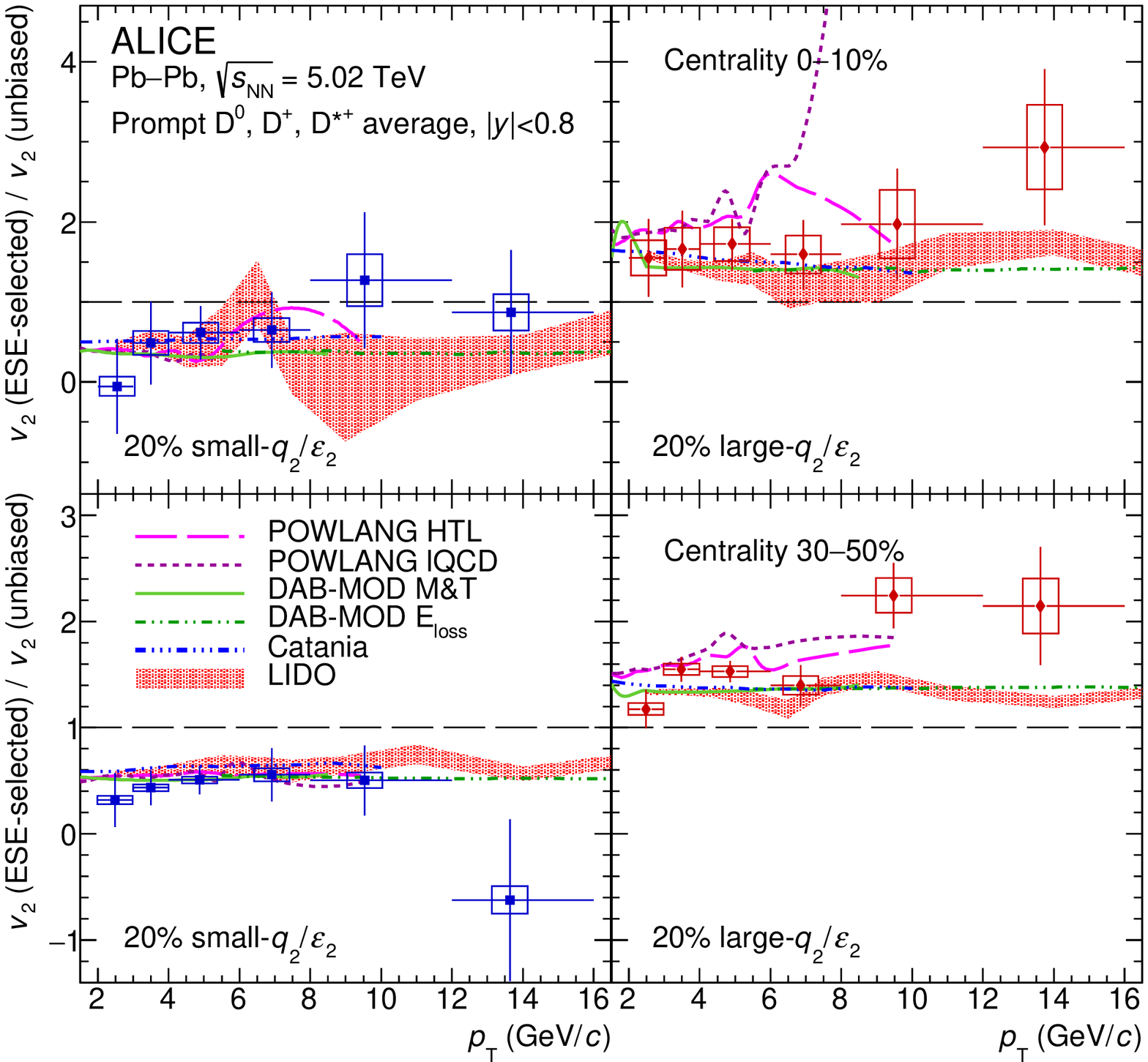}
\caption{Ratio of the average prompt $\Dzero$, $\Dplus$, and $\Dstar$ meson $v_2$ coefficients measured in the small-$q_2$ (left panels) and large-$q_2$ (right panels) selected samples with respect to that of the unbiased sample as a function of $\pt$ in $\PbPb$ collisions at $\sqrtsNN=5.02~\TeV$ for the 0--10\% (top panels) and 30--50\% (bottom panels) centrality classes, compared to model calculations~\cite{Beraudo:2018tpr,Ke:2018tsh,Prado:2016szr,Katz:2019fkc,Plumari:2019hzp}.} 
\label{fig:Daverage_v2Ratio_ESE} 
\end{center}
\end{figure*}

The average $v_2$ of prompt $\Dzero$, $\Dplus$, and $\Dstar$ mesons measured in the ESE-selected samples is shown in Fig.~\ref{fig:Daverage_v2_ESE} for the 0--10\% (top row) and the 30--50\% (bottom row) centrality classes. The measurements in the small-$q_2$ sample are reported in the left column, those in the large-$q_2$ sample in the right column, while the measurements in the unbiased samples recomputed in the same $\pt$ intervals of the ESE analysis are in the middle column. A reduced $\pt$ range ($2<\pt<16~\GeV/c$) and wider $\pt$ intervals compared to the unbiased $v_2$ measurement were adopted due to the limited size of the ESE-selected samples. The average $v_2$ among the three D-meson species was computed as described in Sec.~\ref{sec:results_unb}. In Fig.~\ref{fig:Daverage_v2Ratio_ESE} the ratio between the average D-meson $v_2$ measured in the ESE-selected samples with respect to that in the unbiased sample is depicted. The statistical uncertainties of the ratio were calculated taking into consideration the degree of correlation between the measurements in the ESE-selected and unbiased samples. The systematic uncertainties arising from the centrality dependence of $\Rnflow$, the non-flow contaminations among sub-events, and the correction for the beauty feed-down contribution were considered as fully correlated.  

The D-meson $v_2$ was found to be on average about 50\% higher (lower) in the 20\% of the events with largest (smallest) $q_2$ in both the 0--10\% and 30--50\% centrality classes. No significant centrality dependence was found within the current uncertainties. The corresponding variation of the average $q_2$ in the small-$q_2$ (large-$q_2$) sample with respect to the unbiased one was found to be about 65\% (75\%) and 60\% (65\%) for the 0--10\% and 30--50\% centrality class, respectively. This confirms the correlation between the D-meson azimuthal anisotropy and the collective expansion of the bulk matter already observed in~\cite{Acharya:2018bxo}. This modification of the $v_2$ coefficient was found to be independent of $\pt$ within uncertainties, which might suggest that the ESE selection is related to a global property of the events (i.e. a property that is independent of the measured particle and is related to the entire event). A similar trend was also observed for light-flavour particles~\cite{Adam:2015eta}.

Figures~\ref{fig:Daverage_v2_ESE} and~\ref{fig:Daverage_v2Ratio_ESE} also compare the measured $v_2$ and $v_2$ ratios between ESE-selected and unbiased samples to the \powlang, \lido, \dabmod{}, and \catania{} theoretical predictions. For the \powlang{} model, both the predictions obtained with the transport coefficients from weak coupling (Hard Thermal Loop, HTL~\cite{Alberico:2013bza}) and from lattice QCD calculations (lQCD~\cite{Banerjee:2011ra}) are reported. For the \dabmod{} model, a version based on the heavy-quark transport (M\&T~\cite{Moore:2004tg}) and a parametric model for the heavy-quark energy loss ($E_\mathrm{loss}$~\cite{Betz:2014cza}) were considered. In the \lido{}, \dabmod{}, and \catania{} models the ESE selection is performed with a $q_2$ estimator computed starting from generated quantities~\cite{Ke:2018tsh,Katz:2019fkc,Plumari:2019hzp}, while in the \powlang{} model the elliptic eccentricity $\epsilon_2$ is directly used~\cite{Beraudo:2018tpr}. The $v_2$ measured in the small-$q_2$ sample is described by all the available models within the uncertainties. On the contrary, in the 30--50\% centrality class the \lido{}, \dabmod{}, and \catania{} models underestimate the measurement in the large-$q_2$ sample, which is instead well described by the \powlanghtl{} prediction. In the case of \powlanglqcd, the theoretical prediction is compatible with the measured $v_2$ for $\pt<4~\GeV/c$ and lower for higher $\pt$. The \dabmod{} calculations give a better description of the experimental data with the M\&T approach for $\pt<5~\GeV/c$ and in the $E_\mathrm{loss}$ case for $\pt>5~\GeV/c$. When the ratios between the $v_2$ in the ESE-selected and the unbiased samples are considered, the models seem to better describe the measured values, owing to similar discrepancies between theoretical predictions and experimental data in the ESE-selected and unbiased samples which lead to similar ratio values in the different models. In the small-$q_2$ samples the model predictions are more similar to each other and the discrepancies are less significant, also due to the larger experimental uncertainties.
Interestingly, different implementations of the same model with the studied transport parameterisations (i.e. \powlanghtl{} vs. \powlanglqcd, and \dabmodmet{} vs. \dabmodeloss) give similar predictions, suggesting that the effect of the ESE selection is more related to the initial geometry and the underlying hydrodynamic expansion rather than the dynamic evolution of the heavy quarks in the medium.

\begin{figure*}[!t]
\begin{center}
\includegraphics[width=1.\textwidth]{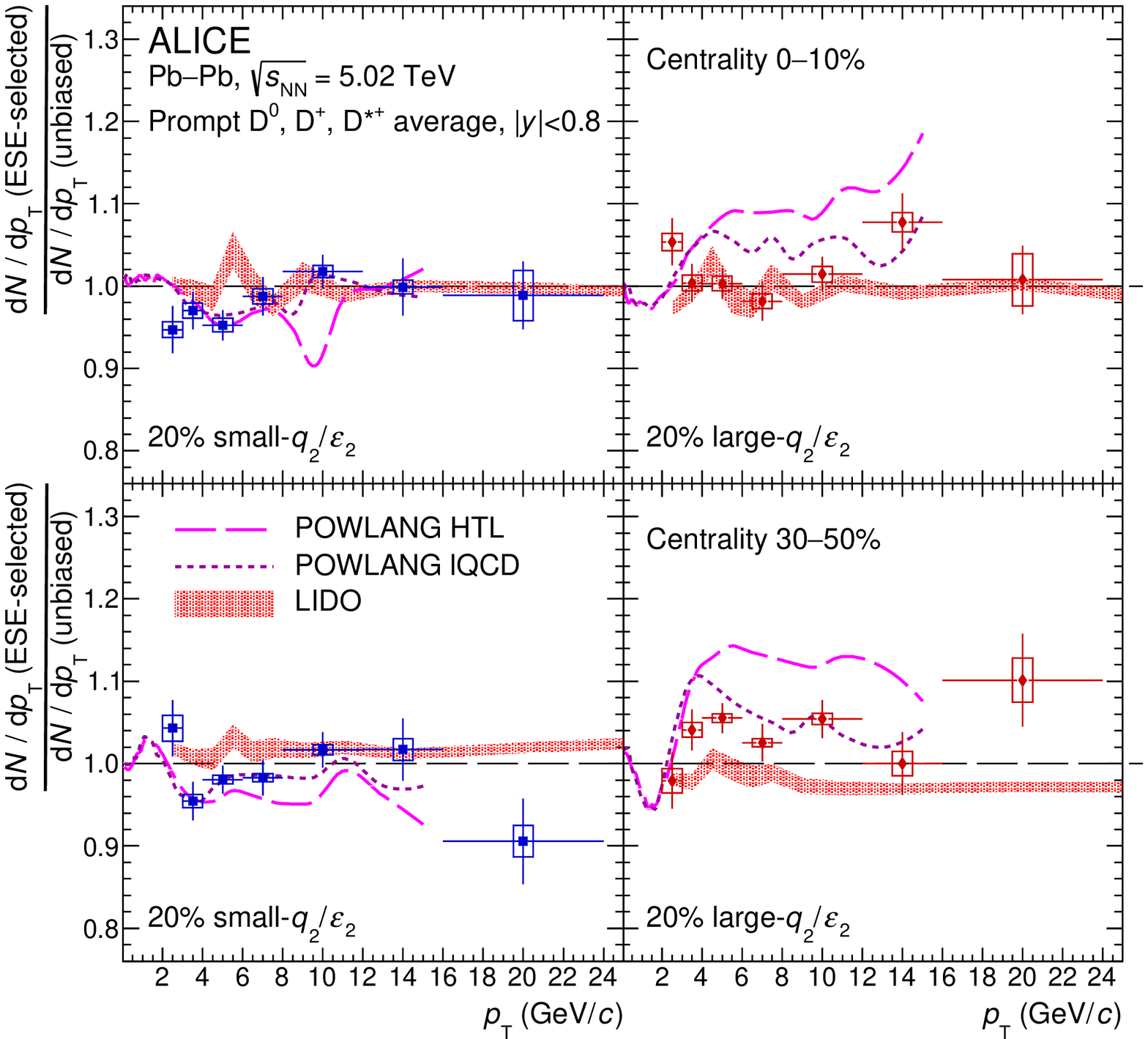}
\caption{Average of the ratio of $\pt$-differential $\Dzero$, $\Dplus$, and $\Dstar$ yields measured in the ESE-selected samples to those in the unbiased sample in $\PbPb$ collisions at $\sqrtsNN=5.02~\TeV$ for the 0--10\% (top panels) and 30--50\% (bottom panels) centrality classes, compared to the \powlang~\cite{Beraudo:2018tpr} and \lido~\cite{Ke:2018tsh,Ke:2018jem} predictions.} 
\label{fig:Daverage_yieldRatio_ESE} 
\end{center}
\end{figure*}

To study a possible interplay between the azimuthal anisotropy of the event and the charm-quark radial flow (at low/intermediate $\pt$) and in-medium energy loss (at high $\pt$), the ratio of the measured per-event yields of prompt $\Dzero$, $\Dplus$, and $\Dstar$ mesons in the ESE-selected and unbiased samples has been calculated as a function of $\pt$ in the range $2<\pt<24~\GeV/c$. The average D-meson ratios, computed by using the inverse of the squared relative statistical uncertainties as weights, are compared to the \powlang{} and \lido{} models in Fig.~\ref{fig:Daverage_yieldRatio_ESE}. The \powlang{} model predicts a hardening (softening) of the $\pt$ distributions in the large (small)-$q_2$ class of events due to an interplay between the radial and elliptic flows, while no significant modification is predicted by the \lido{} model. Within the current precision, the measured per-event yield ratios and are found compatible with unity, and hence to the \lido{} model predictions, and with the \powlang{} model in the case of lQCD, while the measured effect seems to be lower than the effect predicted with HTL transport coefficients.

\section{Conclusions}
The elliptic and triangular flow of $\Dzero$, $\Dplus$, and $\Dstar$ mesons was measured with the SP method at midrapidity ($|y|<0.8$) as a function of $\pt$ in central (0--10\%) and semi-central (30--50\%) Pb--Pb collisions at $\sqrtsNN=5.02~\TeV$. 

Compared to other particle species, the average D-meson $\vn$ harmonics were found to be compatible with the hypothesis of a mass hierarchy for $\pt\lesssim 3~\GeV/c$ as observed for light-flavour hadrons~\cite{Acharya:2018zuq}. At intermediate $\pt$, the D-meson $\vn$ is similar to those of charged pions, lower than those of protons, and higher than those of $\Jpsi$ mesons, supporting the hypothesis of charm-quark hadronisation via coalescence. Moreover, the contribution to the hadronisation of charm quarks from coalescence with light quarks from the medium seems to be necessary in the theoretical models to quantitatively reproduce the measured D-meson $\vn$. For $\pt\gtrsim 8~\GeV/c$, the D-meson $v_2$ and $v_3$ are compatible within uncertainties with the values measured for the other particle species, indicating a similar path-length dependence of the energy loss of high-$\pt$ charm quarks and gluons. The comparison of the measured D-meson $\vn$ with theoretical calculations suggests that the interactions with the hydrodynamically expanding medium impart a positive $v_2$ and $v_3$ to the charm quarks.

The elliptic flow and the modification of the $\pt$ distributions of $\Dzero$, $\Dplus$, and $\Dstar$ mesons were also investigated with the event-shape engineering technique. The D-meson $v_2$ was found to be larger (smaller) in events with larger (smaller) $q_2$, confirming the correlation with average bulk elliptic flow. The ratios of the $\pt$-differential yields measured in the ESE-selected samples and the unbiased sample were found to be compatible with unity. The measurements in the ESE-selected samples are qualitatively described by theoretical calculations and provide new constraints to models based on charm-quark transport in a hydrodynamically expanding medium and charm-quark energy loss in the QGP.

\newenvironment{acknowledgement}{\relax}{\relax}
\begin{acknowledgement}
\section*{Acknowledgements}

The ALICE Collaboration would like to thank all its engineers and technicians for their invaluable contributions to the construction of the experiment and the CERN accelerator teams for the outstanding performance of the LHC complex.
The ALICE Collaboration gratefully acknowledges the resources and support provided by all Grid centres and the Worldwide LHC Computing Grid (WLCG) collaboration.
The ALICE Collaboration acknowledges the following funding agencies for their support in building and running the ALICE detector:
A. I. Alikhanyan National Science Laboratory (Yerevan Physics Institute) Foundation (ANSL), State Committee of Science and World Federation of Scientists (WFS), Armenia;
Austrian Academy of Sciences, Austrian Science Fund (FWF): [M 2467-N36] and Nationalstiftung f\"{u}r Forschung, Technologie und Entwicklung, Austria;
Ministry of Communications and High Technologies, National Nuclear Research Center, Azerbaijan;
Conselho Nacional de Desenvolvimento Cient\'{\i}fico e Tecnol\'{o}gico (CNPq), Financiadora de Estudos e Projetos (Finep), Funda\c{c}\~{a}o de Amparo \`{a} Pesquisa do Estado de S\~{a}o Paulo (FAPESP) and Universidade Federal do Rio Grande do Sul (UFRGS), Brazil;
Ministry of Education of China (MOEC) , Ministry of Science \& Technology of China (MSTC) and National Natural Science Foundation of China (NSFC), China;
Ministry of Science and Education and Croatian Science Foundation, Croatia;
Centro de Aplicaciones Tecnol\'{o}gicas y Desarrollo Nuclear (CEADEN), Cubaenerg\'{\i}a, Cuba;
Ministry of Education, Youth and Sports of the Czech Republic, Czech Republic;
The Danish Council for Independent Research | Natural Sciences, the VILLUM FONDEN and Danish National Research Foundation (DNRF), Denmark;
Helsinki Institute of Physics (HIP), Finland;
Commissariat \`{a} l'Energie Atomique (CEA) and Institut National de Physique Nucl\'{e}aire et de Physique des Particules (IN2P3) and Centre National de la Recherche Scientifique (CNRS), France;
Bundesministerium f\"{u}r Bildung und Forschung (BMBF) and GSI Helmholtzzentrum f\"{u}r Schwerionenforschung GmbH, Germany;
General Secretariat for Research and Technology, Ministry of Education, Research and Religions, Greece;
National Research, Development and Innovation Office, Hungary;
Department of Atomic Energy Government of India (DAE), Department of Science and Technology, Government of India (DST), University Grants Commission, Government of India (UGC) and Council of Scientific and Industrial Research (CSIR), India;
Indonesian Institute of Science, Indonesia;
Centro Fermi - Museo Storico della Fisica e Centro Studi e Ricerche Enrico Fermi and Istituto Nazionale di Fisica Nucleare (INFN), Italy;
Institute for Innovative Science and Technology , Nagasaki Institute of Applied Science (IIST), Japanese Ministry of Education, Culture, Sports, Science and Technology (MEXT) and Japan Society for the Promotion of Science (JSPS) KAKENHI, Japan;
Consejo Nacional de Ciencia (CONACYT) y Tecnolog\'{i}a, through Fondo de Cooperaci\'{o}n Internacional en Ciencia y Tecnolog\'{i}a (FONCICYT) and Direcci\'{o}n General de Asuntos del Personal Academico (DGAPA), Mexico;
Nederlandse Organisatie voor Wetenschappelijk Onderzoek (NWO), Netherlands;
The Research Council of Norway, Norway;
Commission on Science and Technology for Sustainable Development in the South (COMSATS), Pakistan;
Pontificia Universidad Cat\'{o}lica del Per\'{u}, Peru;
Ministry of Science and Higher Education, National Science Centre and WUT ID-UB, Poland;
Korea Institute of Science and Technology Information and National Research Foundation of Korea (NRF), Republic of Korea;
Ministry of Education and Scientific Research, Institute of Atomic Physics and Ministry of Research and Innovation and Institute of Atomic Physics, Romania;
Joint Institute for Nuclear Research (JINR), Ministry of Education and Science of the Russian Federation, National Research Centre Kurchatov Institute, Russian Science Foundation and Russian Foundation for Basic Research, Russia;
Ministry of Education, Science, Research and Sport of the Slovak Republic, Slovakia;
National Research Foundation of South Africa, South Africa;
Swedish Research Council (VR) and Knut \& Alice Wallenberg Foundation (KAW), Sweden;
European Organization for Nuclear Research, Switzerland;
Suranaree University of Technology (SUT), National Science and Technology Development Agency (NSDTA) and Office of the Higher Education Commission under NRU project of Thailand, Thailand;
Turkish Atomic Energy Agency (TAEK), Turkey;
National Academy of  Sciences of Ukraine, Ukraine;
Science and Technology Facilities Council (STFC), United Kingdom;
National Science Foundation of the United States of America (NSF) and United States Department of Energy, Office of Nuclear Physics (DOE NP), United States of America.    
\end{acknowledgement}

\bibliographystyle{utphys}   
\bibliography{dmeson_vn}

\newpage
\appendix
\section{The ALICE Collaboration}
\label{app:collab}

\begingroup
\small
\begin{flushleft}
S.~Acharya\Irefn{org141}\And 
D.~Adamov\'{a}\Irefn{org95}\And 
A.~Adler\Irefn{org74}\And 
J.~Adolfsson\Irefn{org81}\And 
M.M.~Aggarwal\Irefn{org100}\And 
G.~Aglieri Rinella\Irefn{org34}\And 
M.~Agnello\Irefn{org30}\And 
N.~Agrawal\Irefn{org10}\textsuperscript{,}\Irefn{org54}\And 
Z.~Ahammed\Irefn{org141}\And 
S.~Ahmad\Irefn{org16}
% ...
\And
\renewcommand\labelenumi{\textsuperscript{\theenumi}~}

\section*{Affiliation notes}
\renewcommand\theenumi{\roman{enumi}}
\begin{Authlist}
\item \Adef{org*}Deceased
\item \Adef{orgI}Italian National Agency for New Technologies, Energy and Sustainable Economic Development (ENEA), Bologna, Italy
\item \Adef{orgII}Dipartimento DET del Politecnico di Torino, Turin, Italy
\item \Adef{orgIII}M.V. Lomonosov Moscow State University, D.V. Skobeltsyn Institute of Nuclear, Physics, Moscow, Russia
\item \Adef{orgIV}Department of Applied Physics, Aligarh Muslim University, Aligarh, India
\item \Adef{orgV}Institute of Theoretical Physics, University of Wroclaw, Poland
\end{Authlist}

\section*{Collaboration Institutes}
\renewcommand\theenumi{\arabic{enumi}~}
\begin{Authlist}
\item \Idef{org1}A.I. Alikhanyan National Science Laboratory (Yerevan Physics Institute) Foundation, Yerevan, Armenia
\item \Idef{org2}Bogolyubov Institute for Theoretical Physics, National Academy of Sciences of Ukraine, Kiev, Ukraine
\item \Idef{org3}Bose Institute, Department of Physics  and Centre for Astroparticle Physics and Space Science (CAPSS), Kolkata, India
\item \Idef{org4}Budker Institute for Nuclear Physics, Novosibirsk, Russia
\item \Idef{org5}California Polytechnic State University, San Luis Obispo, California, United States
\item \Idef{org6}Central China Normal University, Wuhan, China
\item \Idef{org7}Centre de Calcul de l'IN2P3, Villeurbanne, Lyon, France
\item \Idef{org8}Centro de Aplicaciones Tecnol\'{o}gicas y Desarrollo Nuclear (CEADEN), Havana, Cuba
\item \Idef{org9}Centro de Investigaci\'{o}n y de Estudios Avanzados (CINVESTAV), Mexico City and M\'{e}rida, Mexico
\item \Idef{org10}Centro Fermi - Museo Storico della Fisica e Centro Studi e Ricerche ``Enrico Fermi', Rome, Italy
\item \Idef{org11}Chicago State University, Chicago, Illinois, United States
\item \Idef{org12}China Institute of Atomic Energy, Beijing, China
\item \Idef{org13}Comenius University Bratislava, Faculty of Mathematics, Physics and Informatics, Bratislava, Slovakia
\item \Idef{org14}COMSATS University Islamabad, Islamabad, Pakistan
\item \Idef{org15}Creighton University, Omaha, Nebraska, United States
\item \Idef{org16}Department of Physics, Aligarh Muslim University, Aligarh, India
\item \Idef{org17}Department of Physics, Pusan National University, Pusan, Republic of Korea
\item \Idef{org18}Department of Physics, Sejong University, Seoul, Republic of Korea
\item \Idef{org19}Department of Physics, University of California, Berkeley, California, United States
\item \Idef{org20}Department of Physics, University of Oslo, Oslo, Norway
\item \Idef{org21}Department of Physics and Technology, University of Bergen, Bergen, Norway
\item \Idef{org22}Dipartimento di Fisica dell'Universit\`{a} 'La Sapienza' and Sezione INFN, Rome, Italy
\item \Idef{org23}Dipartimento di Fisica dell'Universit\`{a} and Sezione INFN, Cagliari, Italy
\item \Idef{org24}Dipartimento di Fisica dell'Universit\`{a} and Sezione INFN, Trieste, Italy
\item \Idef{org25}Dipartimento di Fisica dell'Universit\`{a} and Sezione INFN, Turin, Italy
\item \Idef{org26}Dipartimento di Fisica e Astronomia dell'Universit\`{a} and Sezione INFN, Bologna, Italy
\item \Idef{org27}Dipartimento di Fisica e Astronomia dell'Universit\`{a} and Sezione INFN, Catania, Italy
\item \Idef{org28}Dipartimento di Fisica e Astronomia dell'Universit\`{a} and Sezione INFN, Padova, Italy
\item \Idef{org29}Dipartimento di Fisica `E.R.~Caianiello' dell'Universit\`{a} and Gruppo Collegato INFN, Salerno, Italy
\item \Idef{org30}Dipartimento DISAT del Politecnico and Sezione INFN, Turin, Italy
\item \Idef{org31}Dipartimento di Scienze e Innovazione Tecnologica dell'Universit\`{a} del Piemonte Orientale and INFN Sezione di Torino, Alessandria, Italy
\item \Idef{org32}Dipartimento di Scienze MIFT, Universit\`{a} di Messina, Messina, Italy
\item \Idef{org33}Dipartimento Interateneo di Fisica `M.~Merlin' and Sezione INFN, Bari, Italy
\item \Idef{org34}European Organization for Nuclear Research (CERN), Geneva, Switzerland
\item \Idef{org35}Faculty of Electrical Engineering, Mechanical Engineering and Naval Architecture, University of Split, Split, Croatia
\item \Idef{org36}Faculty of Engineering and Science, Western Norway University of Applied Sciences, Bergen, Norway
\item \Idef{org37}Faculty of Nuclear Sciences and Physical Engineering, Czech Technical University in Prague, Prague, Czech Republic
\item \Idef{org38}Faculty of Science, P.J.~\v{S}af\'{a}rik University, Ko\v{s}ice, Slovakia
\item \Idef{org39}Frankfurt Institute for Advanced Studies, Johann Wolfgang Goethe-Universit\"{a}t Frankfurt, Frankfurt, Germany
\item \Idef{org40}Fudan University, Shanghai, China
\item \Idef{org41}Gangneung-Wonju National University, Gangneung, Republic of Korea
\item \Idef{org42}Gauhati University, Department of Physics, Guwahati, India
\item \Idef{org43}Helmholtz-Institut f\"{u}r Strahlen- und Kernphysik, Rheinische Friedrich-Wilhelms-Universit\"{a}t Bonn, Bonn, Germany
\item \Idef{org44}Helsinki Institute of Physics (HIP), Helsinki, Finland
\item \Idef{org45}High Energy Physics Group,  Universidad Aut\'{o}noma de Puebla, Puebla, Mexico
\item \Idef{org46}Hiroshima University, Hiroshima, Japan
\item \Idef{org47}Hochschule Worms, Zentrum  f\"{u}r Technologietransfer und Telekommunikation (ZTT), Worms, Germany
\item \Idef{org48}Horia Hulubei National Institute of Physics and Nuclear Engineering, Bucharest, Romania
\item \Idef{org49}Indian Institute of Technology Bombay (IIT), Mumbai, India
\item \Idef{org50}Indian Institute of Technology Indore, Indore, India
\item \Idef{org51}Indonesian Institute of Sciences, Jakarta, Indonesia
\item \Idef{org52}INFN, Laboratori Nazionali di Frascati, Frascati, Italy
\item \Idef{org53}INFN, Sezione di Bari, Bari, Italy
\item \Idef{org54}INFN, Sezione di Bologna, Bologna, Italy
\item \Idef{org55}INFN, Sezione di Cagliari, Cagliari, Italy
\item \Idef{org56}INFN, Sezione di Catania, Catania, Italy
\item \Idef{org57}INFN, Sezione di Padova, Padova, Italy
\item \Idef{org58}INFN, Sezione di Roma, Rome, Italy
\item \Idef{org59}INFN, Sezione di Torino, Turin, Italy
\item \Idef{org60}INFN, Sezione di Trieste, Trieste, Italy
\item \Idef{org61}Inha University, Incheon, Republic of Korea
\item \Idef{org62}Institute for Nuclear Research, Academy of Sciences, Moscow, Russia
\item \Idef{org63}Institute for Subatomic Physics, Utrecht University/Nikhef, Utrecht, Netherlands
\item \Idef{org64}Institute of Experimental Physics, Slovak Academy of Sciences, Ko\v{s}ice, Slovakia
\item \Idef{org65}Institute of Physics, Homi Bhabha National Institute, Bhubaneswar, India
\item \Idef{org66}Institute of Physics of the Czech Academy of Sciences, Prague, Czech Republic
\item \Idef{org67}Institute of Space Science (ISS), Bucharest, Romania
\item \Idef{org68}Institut f\"{u}r Kernphysik, Johann Wolfgang Goethe-Universit\"{a}t Frankfurt, Frankfurt, Germany
\item \Idef{org69}Instituto de Ciencias Nucleares, Universidad Nacional Aut\'{o}noma de M\'{e}xico, Mexico City, Mexico
\item \Idef{org70}Instituto de F\'{i}sica, Universidade Federal do Rio Grande do Sul (UFRGS), Porto Alegre, Brazil
\item \Idef{org71}Instituto de F\'{\i}sica, Universidad Nacional Aut\'{o}noma de M\'{e}xico, Mexico City, Mexico
\item \Idef{org72}iThemba LABS, National Research Foundation, Somerset West, South Africa
\item \Idef{org73}Jeonbuk National University, Jeonju, Republic of Korea
\item \Idef{org74}Johann-Wolfgang-Goethe Universit\"{a}t Frankfurt Institut f\"{u}r Informatik, Fachbereich Informatik und Mathematik, Frankfurt, Germany
\item \Idef{org75}Joint Institute for Nuclear Research (JINR), Dubna, Russia
\item \Idef{org76}Korea Institute of Science and Technology Information, Daejeon, Republic of Korea
\item \Idef{org77}KTO Karatay University, Konya, Turkey
\item \Idef{org78}Laboratoire de Physique des 2 Infinis, Ir\`{e}ne Joliot-Curie, Orsay, France
\item \Idef{org79}Laboratoire de Physique Subatomique et de Cosmologie, Universit\'{e} Grenoble-Alpes, CNRS-IN2P3, Grenoble, France
\item \Idef{org80}Lawrence Berkeley National Laboratory, Berkeley, California, United States
\item \Idef{org81}Lund University Department of Physics, Division of Particle Physics, Lund, Sweden
\item \Idef{org82}Nagasaki Institute of Applied Science, Nagasaki, Japan
\item \Idef{org83}Nara Women{'}s University (NWU), Nara, Japan
\item \Idef{org84}National and Kapodistrian University of Athens, School of Science, Department of Physics , Athens, Greece
\item \Idef{org85}National Centre for Nuclear Research, Warsaw, Poland
\item \Idef{org86}National Institute of Science Education and Research, Homi Bhabha National Institute, Jatni, India
\item \Idef{org87}National Nuclear Research Center, Baku, Azerbaijan
\item \Idef{org88}National Research Centre Kurchatov Institute, Moscow, Russia
\item \Idef{org89}Niels Bohr Institute, University of Copenhagen, Copenhagen, Denmark
\item \Idef{org90}Nikhef, National institute for subatomic physics, Amsterdam, Netherlands
\item \Idef{org91}NRC Kurchatov Institute IHEP, Protvino, Russia
\item \Idef{org92}NRC \guillemotleft Kurchatov\guillemotright~Institute - ITEP, Moscow, Russia
\item \Idef{org93}NRNU Moscow Engineering Physics Institute, Moscow, Russia
\item \Idef{org94}Nuclear Physics Group, STFC Daresbury Laboratory, Daresbury, United Kingdom
\item \Idef{org95}Nuclear Physics Institute of the Czech Academy of Sciences, \v{R}e\v{z} u Prahy, Czech Republic
\item \Idef{org96}Oak Ridge National Laboratory, Oak Ridge, Tennessee, United States
\item \Idef{org97}Ohio State University, Columbus, Ohio, United States
\item \Idef{org98}Petersburg Nuclear Physics Institute, Gatchina, Russia
\item \Idef{org99}Physics department, Faculty of science, University of Zagreb, Zagreb, Croatia
\item \Idef{org100}Physics Department, Panjab University, Chandigarh, India
\item \Idef{org101}Physics Department, University of Jammu, Jammu, India
\item \Idef{org102}Physics Department, University of Rajasthan, Jaipur, India
\item \Idef{org103}Physikalisches Institut, Eberhard-Karls-Universit\"{a}t T\"{u}bingen, T\"{u}bingen, Germany
\item \Idef{org104}Physikalisches Institut, Ruprecht-Karls-Universit\"{a}t Heidelberg, Heidelberg, Germany
\item \Idef{org105}Physik Department, Technische Universit\"{a}t M\"{u}nchen, Munich, Germany
\item \Idef{org106}Politecnico di Bari, Bari, Italy
\item \Idef{org107}Research Division and ExtreMe Matter Institute EMMI, GSI Helmholtzzentrum f\"ur Schwerionenforschung GmbH, Darmstadt, Germany
\item \Idef{org108}Rudjer Bo\v{s}kovi\'{c} Institute, Zagreb, Croatia
\item \Idef{org109}Russian Federal Nuclear Center (VNIIEF), Sarov, Russia
\item \Idef{org110}Saha Institute of Nuclear Physics, Homi Bhabha National Institute, Kolkata, India
\item \Idef{org111}School of Physics and Astronomy, University of Birmingham, Birmingham, United Kingdom
\item \Idef{org112}Secci\'{o}n F\'{\i}sica, Departamento de Ciencias, Pontificia Universidad Cat\'{o}lica del Per\'{u}, Lima, Peru
\item \Idef{org113}St. Petersburg State University, St. Petersburg, Russia
\item \Idef{org114}Stefan Meyer Institut f\"{u}r Subatomare Physik (SMI), Vienna, Austria
\item \Idef{org115}SUBATECH, IMT Atlantique, Universit\'{e} de Nantes, CNRS-IN2P3, Nantes, France
\item \Idef{org116}Suranaree University of Technology, Nakhon Ratchasima, Thailand
\item \Idef{org117}Technical University of Ko\v{s}ice, Ko\v{s}ice, Slovakia
\item \Idef{org118}The Henryk Niewodniczanski Institute of Nuclear Physics, Polish Academy of Sciences, Cracow, Poland
\item \Idef{org119}The University of Texas at Austin, Austin, Texas, United States
\item \Idef{org120}Universidad Aut\'{o}noma de Sinaloa, Culiac\'{a}n, Mexico
\item \Idef{org121}Universidade de S\~{a}o Paulo (USP), S\~{a}o Paulo, Brazil
\item \Idef{org122}Universidade Estadual de Campinas (UNICAMP), Campinas, Brazil
\item \Idef{org123}Universidade Federal do ABC, Santo Andre, Brazil
\item \Idef{org124}University of Cape Town, Cape Town, South Africa
\item \Idef{org125}University of Houston, Houston, Texas, United States
\item \Idef{org126}University of Jyv\"{a}skyl\"{a}, Jyv\"{a}skyl\"{a}, Finland
\item \Idef{org127}University of Liverpool, Liverpool, United Kingdom
\item \Idef{org128}University of Science and Technology of China, Hefei, China
\item \Idef{org129}University of South-Eastern Norway, Tonsberg, Norway
\item \Idef{org130}University of Tennessee, Knoxville, Tennessee, United States
\item \Idef{org131}University of the Witwatersrand, Johannesburg, South Africa
\item \Idef{org132}University of Tokyo, Tokyo, Japan
\item \Idef{org133}University of Tsukuba, Tsukuba, Japan
\item \Idef{org134}Universit\'{e} Clermont Auvergne, CNRS/IN2P3, LPC, Clermont-Ferrand, France
\item \Idef{org135}Universit\'{e} de Lyon, Universit\'{e} Lyon 1, CNRS/IN2P3, IPN-Lyon, Villeurbanne, Lyon, France
\item \Idef{org136}Universit\'{e} de Strasbourg, CNRS, IPHC UMR 7178, F-67000 Strasbourg, France, Strasbourg, France
\item \Idef{org137}Universit\'{e} Paris-Saclay Centre d'Etudes de Saclay (CEA), IRFU, D\'{e}partment de Physique Nucl\'{e}aire (DPhN), Saclay, France
\item \Idef{org138}Universit\`{a} degli Studi di Foggia, Foggia, Italy
\item \Idef{org139}Universit\`{a} degli Studi di Pavia, Pavia, Italy
\item \Idef{org140}Universit\`{a} di Brescia, Brescia, Italy
\item \Idef{org141}Variable Energy Cyclotron Centre, Homi Bhabha National Institute, Kolkata, India
\item \Idef{org142}Warsaw University of Technology, Warsaw, Poland
\item \Idef{org143}Wayne State University, Detroit, Michigan, United States
\item \Idef{org144}Westf\"{a}lische Wilhelms-Universit\"{a}t M\"{u}nster, Institut f\"{u}r Kernphysik, M\"{u}nster, Germany
\item \Idef{org145}Wigner Research Centre for Physics, Budapest, Hungary
\item \Idef{org146}Yale University, New Haven, Connecticut, United States
\item \Idef{org147}Yonsei University, Seoul, Republic of Korea
\end{Authlist}
\endgroup
\end{document}